\documentclass{llncs}
\usepackage{amssymb}

\newtheorem{Proposition}{Proposition}
\newtheorem{Fact}{Fact}
\newtheorem{Definition}{Definition}

\usepackage{amsmath}
\usepackage{amsfonts}
\usepackage{amssymb}

\usepackage{braket}

\newcommand{\FF}{\mathbb{F}}
\newcommand{\Lin}{\mathrm{Lin}}

\frontmatter
\mainmatter     
\title{On weakly APN functions and 4-bit S-boxes}
\author{Claudio Fontanari\and Valentina Pulice\and Anna Rimoldi\and Massimiliano Sala}
\institute{Department of Mathematics, University of Trento, Italy,\\
\email{fontanar@science.unitn.it}\\
\email{pulice@science.unitn.it}\\
\email{maxsalacodes@gmail.com}\\ 
eRISCS, Universite de la Mediterranee, Marseille, France,\\
\email{rimoldi@science.unitn.it}\\
}

\date{}
\begin{document}
\maketitle

\begin{abstract}
S-Boxes are important security components of block ciphers.
We provide theoretical results on necessary or sufficient criteria
for an (invertible) 4-bit S-Box to be weakly APN.
Thanks to a classification of 4-bit invertible S-Boxes achieved independently 
by De Canni\'ere and Leander-Poschmann, we can strengthen our
results with a computer-aided proof.
We also propose a class of 4-bit S-Boxes which are very strong from a security point of view.
\end{abstract}


\section{Introduction}

We consider block ciphers acting on a vector space $(\mathbb{F}_2)^n$. 
It is important to identify conditions on the components of the cipher that
may ensure its security. There are many competing notions of security, 
hence several kinds of security criteria, and some of them focus on the role 
of the S-Boxes. For a large class of nowadays block ciphers, the S-Boxes are 
bijective vectorial Boolean functions $f: (\mathbb{F}_2)^m \to (\mathbb{F}_2)^m$, hence they are functions from the finite field $(\mathbb{F}_2)^m$ to itself.

In this paper we focus on $4$-bit S-Boxes, as used for example in SERPENT (\cite{serpent}) and PRESENT (\cite{present}), 
although we present also a theorem for the general case.
Several security criteria are affine-invariant and this justifies
the work done to achieve the classification of 4-bit S-Boxes
in affine-equivalence classes, as done for example by
De Canni\'ere (\cite{dec}) and Leander and Poschmann (\cite{Sboxes})
(these classifications have been achieved independently).

There is a new security criteria for S-Boxes which is affine-invariant, the weakly differential uniformity.
Particularly interesting is the concept of weakly APN.
We determine several conditions (some computational and some theoretical), 
which are either sufficient or necessary for a $4$-bit vectorial Boolean function 
to be weakly APN.

Our paper is structured as follows.
In Section. \ref{theoretical}, we introduce and motivate the notion 
of \emph{weakly APN function}, highlighting the case of dimension $4$.
In Section. \ref{secth} we present our theoretical results,
including a theorem for any dimension.
In Section. \ref{computational} we discuss our computational results.
Finally, in Section. \ref{concl} we provide further computations that may be interesting and we draw our conclusions.

\section{Preliminaries on weakly APN functions}\label{theoretical}

Without loss of generality, in the sequel we consider only Boolean functions 
$f: (\mathbb{F}_2)^m \to (\mathbb{F}_2)^m$ such that $f(0)=0$.
We also write  $\hat{f}_u(x) := f(x+u)+f(x)$ (the \emph{derivative} of $f$) and 
$\mathrm{Im}(f)=\{ f(x) \mid x\in (\mathbb{F}_2)^m\}$ (the \emph{image} of $f$).

A notion of non-linearity for S-Boxes that has received a lot of attention is
the following.

\begin{definition}
The function $f$ is \emph{$\delta$-differentially uniform} if for any $u \in (\mathbb{F}_2)^m \setminus \{ 0 \}$ and 
for any $v \in  (\mathbb{F}_2)^m$, $\vert \{ x \in  (\mathbb{F}_2)^m: 
\hat{f}_u(x) = v \} \vert \le \delta$ .

If $f$ is $2$-differentially uniform, then it is called an \emph{Almost Perfectly Nonlinear (APN)} function.
\end{definition}
The property of being $\delta$-differentially uniform is an affine-invariant.
W.r.t. diffentially uniformity, the best S-Boxes are the APN S-Boxes.
APN functions are indeed a very hot research topic 
(see for instance the recent contributions \cite{aubry} and \cite{bracken}).
Unfortunately, for some even dimensions, no APN permutation exists.
This is the case for dimension $m=4$, which has cryptographic significance
at least for SERPENT and PRESENT.
In this case, the best we can have is $\delta=4$.

There is a natural generalization of differential uniformity presented
recently in \cite{CDS}, which we recall in the following definition.
\begin{definition}
The function $f$ is  \emph{weakly $\delta$-differentially uniform} if for any 
$u \in (\mathbb{F}_2)^m \setminus \{ 0 \}$ we have $\vert \mathrm{Im}(\hat{f}_u) \vert > 2^{m-1}/\delta$. 

If $f$ is weakly $2$-differentially uniform, then it is called a \emph{weakly Almost Perfectly Nonlinear (weakly APN)} function.
\end{definition}
By \cite{CDS}, \S 4, Fact 3, a $\delta$-differentially uniform map is weakly $\delta$-differentially uniform, and is easy to check that 
weak $\delta$-differential uniformity is affine-invariant. 

The significance for the previous definition lies in \cite{CDS}, Theorem 4.4.
To appreciate it we need another definition.
\begin{definition}
A function $f$ is \emph{strongly $l$-anti-invariant} if for 
any two subspaces $V,W \leq (\mathbb{F}_2)^m$ such that $f(V)=W$ then either $\dim(V)=\dim(W)<m-l$ or 
$V=W=(\mathbb{F}_2)^m$. 
\end{definition}

An iterated block cipher is obtained by the composition of several rounds (or round functions),  
i.e., key-dependent permutations of the message/cipher space. To avoid potential weaknesses 
of a given cipher $\mathcal{C}$, it is desirable that the permutation group $\Gamma_\infty(\mathcal{C})$ 
generated by its round functions with the key varying in the key space is primitive (for instance, a way 
to construct a trapdoor using imprimitivity is presented in \cite{trapdoor}).
Translation-based ciphers (see \cite{CDS}, Def. 3.1) form an interesting class of iterated block ciphers
containing AES\cite{AES}, SERPENT, PRESENT. According to Theorem.~4.4 in \cite{CDS},
if $\mathcal{C}$ is a translation-based cipher and each brick $\gamma'$ of every parallel S-Box $\gamma$ used in 
the proper round under consideration is both weakly $2^r$-differentially uniform and strongly $r$-anti-invariant
for some $r$ with $1 \le r \le m/2$, then $\Gamma_\infty(\mathcal{C})$ is primitive. 
It may seem that Theorem~4.4 in \cite{CDS} requires too strong conditions in order to ensure primitivity, but 
indeed they turn out to be quite natural, as shown in \cite{CDS}, \S 5. 
In the case of $4$-bit S-Boxes, we have only two possibilities: $r=1$, requiring every $\gamma'$ 
to be both strongly $1$-anti-invariant (which always holds if it is maximally non-linear, see for instance 
\cite{CDS}, footnote 4 on p. 347) and weakly APN; 
or $r=2$, requiring every $\gamma'$ to be both weakly $4$-differentially uniform (which always holds if it is $4$-differentially uniform) 
and $2$-strongly-anti-invariant.

\section{Theoretical results on weakly APN functions}
\label{secth}

Our first result is to show that for $4$-differentially uniform functions 
the case $r=2$ of Theorem~4.4 in \cite{CDS} is just a sub-case of the case
$r=1$.

\begin{Proposition}\label{invariant}
Let $f: (\mathbb{F}_2)^4 \to (\mathbb{F}_2)^4$ be a Boolean function such that 

(i) $f$ is $4$-differentially uniform

(ii) $f$ is strongly $2$-anti-invariant.

\noindent Then $f$ is weakly APN.   
\end{Proposition} 

\proof Assume by contradiction that $\vert \mathrm{Im}(\hat{f}_u) \vert \le 4$. 
Then from (i) we deduce that $\vert \hat{f}_u^{-1}(y) \vert = 4$ for every $y\in \mathrm{Im}(\hat{f}_u)$. Hence we have 
$\hat{f}_u^{-1}(f(u)) = \{0,u,x,u+x \}$ for some $x$, in particular $\hat{f}_u^{-1}(f(u))$ is a $2$-dimensional vector 
subspace. On the other hand, $\hat{f}_u(x)=\hat{f}_u(u)$ implies $f(x+u)=f(u)-f(x)$. It follows that $f(\{0,u,x,u+x \})$ 
is a $2$-dimensional vector subspace, contradicting~(ii). 

\qed   

In other words, Proposition \ref{invariant} provides some sufficient
conditions for a 4-bit S-Box to be weakly APN.
Other sufficient conditions are presented in the next proposition
and are based on the following non-linearity measures:
\begin{equation}\label{degreecomp}
 n_i(f)=\vert \{v\in(\FF_2)^m\setminus\{0\}: \deg(<f,v>)=i\} \vert
\end{equation}
 and
\begin{equation}\label{constant}
 \hat{n}(f)=\max_{u\in(\FF_2)^m\setminus{\{0\}}}{\vert\{v\in(\FF_2)^m\setminus{\{0\}}:\deg(<\hat{f}_u,v>)=0\}\vert}\,.
\end{equation}

\begin{Proposition}\label{derivative}
Let $f: (\mathbb{F}_2)^4 \to (\mathbb{F}_2)^4$ be a Boolean function such that 
$\hat{n}(f)=0$.

\noindent Then $f$ is weakly APN.   
\end{Proposition}

\proof 
Let $(\mathbb{F}_2)^4 = \{ x_1, \ldots, x_{16} \}$ and given $u\in(\FF_2)^m\setminus{\{0\}}$ let $M = (m_{ij}) \in (\mathbb{F}_2)^{4 \times 16}$ 
with $m_{ij} := (\hat{f}_u)_i(x_j)$. By definition, $f$ is weakly APN if and only if 
$\vert \mathrm{Im}(\hat{f}_u) \vert > 4$, hence if and only if $M$ has more than $4$ distinct 
columns. 

Assume by contradiction that $M$ has $n \le 4$ distinct columns and let $M' \in 
(\mathbb{F}_2)^{4 \times n}$ be the corresponding submatrix. 

If $M'$ has rank $4$, then we may write $(1,1,1,1)$ as a linear combination of the rows of $M'$:
$$
(1,1,1,1) = a M'_1 + b M'_2 + c M'_3 + d M'_4.
$$
Since all the other columns of $M$ are equal to the columns of $M'$, we may write $(1,\ldots,1) \in 
(\mathbb{F}_2)^{16}$ as the same linear combination of the rows of $M$: 
$$
(1,\ldots,1) = a M_1 + b M_2 + c M_3 + d M_4.
$$
Hence the function $< \hat{f}_u,(a,b,c,d) >$ is the constant $1$, contradiction. 

If instead $M'$ has rank $\le 3$, then we may write $(0,0,0,0)$ as a nonzero linear combination of the rows of $M'$:
$$
(0,0,0,0) = a M'_1 + b M'_2 + c M'_3 + d M'_4.
$$
Since all the other columns of $M$ are equal to the columns of $M'$, we may write $(0,\ldots,0) \in 
(\mathbb{F}_2)^{16}$ as the same linear combination of the rows of $M$: 
$$
(0,\ldots,0) = a M_1 + b M_2 + c M_3 + d M_4.
$$
Hence the function $< \hat{f}_u,(a,b,c,d) >$ is the constant $0$, contradiction. 

\qed 

The following partial converse to Proposition \ref{derivative} gives
necessary conditions and holds for \emph{any} $m\ge 2$.
\begin{theorem}\label{converse}
Let $f: (\mathbb{F}_2)^m \to (\mathbb{F}_2)^m$ be a (weakly) APN function. 

\noindent Then
$\hat{n}(f)\leq 1$.
\end{theorem}

\proof Let $f = (f_1,f_2,\ldots,f_m)$ with $f_i: (\mathbb{F}_2)^m \to \mathbb{F}_2$
and assume by contradiction that both $< \hat{f}_u,v_1 >$ and $< \hat{f}_u,v_2 >$ are constant for some 
$u, v_1 \ne v_2 \in (\mathbb{F}_2)^m \setminus \{ 0 \}$. Up to a linear transformation sending $v_1$ to 
$(1,0,0,\ldots,0)$ and $v_2$ to $(0,1,0,\ldots,0)$, without loss of generality we may assume that both 
$(\hat{f_u})_1 = \hat{(f_1)}_u$ and $(\hat{f_u})_2 = \hat{(f_2)}_u$ are constant. 
It follows that $\vert \mathrm{Im}(\hat{f}_u) \vert \le 2^{m-2}$ and $f$ is not weakly APN, 
contradiction.

\qed

As an application of Theorem \ref{converse}, we obtain the following:

\begin{Proposition}\label{degree}
Let $f: (\mathbb{F}_2)^4 \to (\mathbb{F}_2)^4$ be a weakly APN permutation.  

\noindent Then $\deg(f) = 3$ and $n_3(f) \in \{12,14,15 \}$. 
\end{Proposition}

\proof It is well-known that $\deg{f} \le 3$ (see for instance \cite{permutation}).
If 
$$
  \vert \{v \in (\mathbb{F}_2)^4 \setminus \{ 0 \}: \deg (<f,v>) \le 2 \} \vert \le 5
$$ 
then our claim holds, 
since $\{v \in (\mathbb{F}_2)^4 \setminus \{ 0 \}: \deg (<f,v>) \le 2 \} \cup \{0 \}$ is a vector subspace of  
$(\mathbb{F}_2)^4$.

 Let $f = (f_1,f_2,f_3,f_4)$ with $f_i: (\mathbb{F}_2)^4 \to \mathbb{F}_2$ and assume by contradiction 
that $\deg(S) \le 2$ for $6$ different linear combinations $S = \sum_{i=1}^4 v_i f_i$. From the basic theory of quadratic 
Boolean functions (see for instance \cite{quadratic}, \S 2.2), it follows that the derivative $\hat{S}_u$ is constant 
for every $u \in V(S) \subseteq (\mathbb{F}_2)^4$, where $V(S)$ is a vector subspace of dimension $0$ if and only if $S$ 
is bent, $4$ if and only if $S$ is linear (affine), and $2$ otherwise. Now, $S$ is not bent since it is balanced (see 
for instance \cite{balanced}) and bent functions are never balanced (see for instance \cite{bent}). 
Thus $\dim V(S) \ge 2$ for every $S$ and $\vert V(S) \setminus \{ 0 \} \vert \ge 3$, in particular $6$ sets 
$V(S) \setminus \{ 0 \} \subseteq (\mathbb{F}_2)^4 \setminus \{ 0 \}$ cannot be disjoint. 
Hence there is $u \in (\mathbb{F}_2)^4 \setminus \{ 0 \}$ and two different non-zero linear combinations $S_1$ and $S_2$ 
such that both $\hat{(S_1)}_u$ and $\hat{(S_2)}_u$ are constant and this contradicts Theorem~\ref{converse}.

\qed  

\section{Computational results on weakly APN function}
\label{computational}

The problem of classifying (invertible) S-Boxes $f:(\FF_2)^m\to(\FF_2)^m$ 
(w.r.t. affine-equivalence) was solved in \cite{dec,Sboxes}
in the case $m=4$   and has been recently checked in
\cite{Pul,saarinen}.
%
%
%
By a direct check on the class representatives, we may draw 
a series of consequences, that we call \emph{Facts}.

First of all, we see that three of our theoretical results cannot be
inverted, as follows.
\begin{Fact}\label{no-converse-invariant}
The converse of Proposition \ref{invariant} does not hold.
\end{Fact}
\proof
$(0,1,2,13,4,15,14,7,8,3,5,9,10,6,12,11)$ is weakly APN but is \emph{not}
4-differentially uniform.
\qed
\begin{Fact}\label{no-converse-derivative}
The converse of Proposition \ref{derivative} does not hold. 
\end{Fact}
\proof 
$(0,1,2,13,4,15,14,7,8,3,5,9,10,6,12,11)$ is weakly APN but $\hat{n}=1$.
\qed
\begin{Fact}\label{no-converse-converse}
The converse of Theorem \ref{converse} does not hold.
\end{Fact}
\proof 
For $f=(0,1,2,7,4,10,15,9,8,3,13,14,12,5,6,11)$ we have $\hat{n}(f)=1$
but $f$ is not weakly APN.
\qed

Next, we can strengthen Proposition \ref{degree}:
\begin{Fact}\label{refine-degree}
Let $f: (\mathbb{F}_2)^4 \to (\mathbb{F}_2)^4$ be a weakly APN permutation.  
\noindent Then $\deg(f) = 3$ and $n_3(f) \in \{14,15 \}$. 
\end{Fact}
Unfortunately, the previous fact cannot be inverted:
\begin{Fact}
The converse of Fact \ref{refine-degree} does not hold.
\end{Fact}
\proof 
For $f=(0,1,2,7,4,10,15,9,8,3,13,14,12,5,6,11)$
 we have $\deg(f) = 3$ and  $n_3(f)=14$,
but $f$ is not weakly APN.
\qed

Finally, we want to provide some sufficient conditions 
(for $f$ to be weakly APN), involving also the following classical concept of non-linearity:
\begin{definition}
$$
  \mathrm{Lin}(f)= \max_{a\in (\FF_2)^m,\,b\in(\FF_2)^m\setminus\{0\}}{\vert <f,b>^{\mathcal{W}}(a)}\vert\,,
$$
where $\mathcal{W}$ denotes the Walsh coefficient (see for instance (1) in 
\cite{Sboxes}).
\end{definition}
Since for $m=4$ we have that the best $f$'s have $\Lin(f)=8$, we find
of interest our following result:
\begin{Fact}\label{sufficient}
 Let $f:(\FF_2)^4\to(\FF_2)^4$ be a Boolean permutation such that
$$
   \mathrm{Lin}(f)=8, \quad f \mbox{ is }4-\mbox{differentially uniform}, \quad n_3(f)\ge 14 \,.
$$
\noindent Then $f$ is weakly APN.
\end{Fact}
%
Regrettably,  the assumptions of Fact \ref{sufficient} cannot be weakened.
We provide two (affine-independent) counterexamples:
\begin{itemize}
\item with $f=(0,1,2,12,4,13,11,10,8,15,5,9,6,14,7,3)$ we have
     $\Lin(f)=8$ and $n_3(f)= 14$, but $f$ is not weakly APN,
\item with $f=(0,1,2,12,4,6,14,5,8,3,13,10,9,7,15,11)$ we have that
     $f$ is \\
     $4$-differentially uniform and that $n_3(f)= 14$, but again
     $f$ is not weakly APN.
\end{itemize}

\section{More computational results and conclusions}
\label{concl}

Let we recall from \cite{Sboxes} the further measures of non-linearity:
\begin{itemize}
 \item [-] $
\mathrm{Lin}_1(f)=\max_{\substack{a,b\in(\FF_2)^m\\ \mathrm{w}(a)=\mathrm{w}(b)=1}}{\{\vert{<f,b>^{\mathcal{W}}(a)}\}\vert}\,$,
\item[-] $\mathrm{Diff}_1(f)=\max_{\substack{a,b\in(\FF_2)^m\\ \mathrm{w}(a)=\mathrm{w}(b)=1}}{\{\vert{\hat{f}_a}^{-1}(b)\vert \}}\,$.
\end{itemize}
Then we introduce a new class of S-Boxes suitable for block ciphers construction:
\begin{Definition}
 We say that a Boolean permutation $f:(\FF_2)^4\to(\FF_2)^4$ is a \emph{strong} S-Box if $f$ is weakly APN, $4$-differentially uniform, and
$$
   \mathrm{Lin}(f)=8,  \quad \mathrm{Diff}_1(f)=0
   \quad \mathrm{Lin}_1(f)=4, \quad n_3(f)\ge 14 \,.
$$
Morever, we say that $f$ is \emph{very strong} if it is strong and strongly $2$-anti-invariant.
\end{Definition}
Note that a very strong function is in particular both optimal
(\cite{Sboxes}, Def. 1) and Serpent-type (\cite{Sboxes}, Def. 2),
and also it satisfies Theorem. 4.4 of \cite{CDS}.
%
%
%
A direct computation (see \cite{Pul}) allows us to conclude:
\begin{Fact} \label{number}
 There are $55296$ strong S-Boxes and $2304$ very strong ones.
\end{Fact}
\begin{remark}
As in the rest of the paper, all statements in this section assume $f(0)=0$. So Fact \ref{number} implies that there are
actually $55296*16=884736$ invertible 4-bit S-Boxes equivalent via a translation to strong S-Boxes, therefore sharing their security 
robustness. The same goes for $2304*16=36864$ S-Boxes equivalent to very strong S-Boxes.
\end{remark}

Following \cite{Sboxes}, we have tested the properties of the S-Boxes used in SERPENT, denoted by $S_0,S_1,\ldots,S_7$ (for details see \cite{Pul}), and we get:
\begin{Fact}
The S-Boxes $S_3,S_4,S_5,S_7$ are strong. None of the $S_i$'s is very strong.
\end{Fact}

In conclusion, we have considered the link between the recent notion of weakly APN function and several more traditional non-linearity properties, such as differential uniformity, algebraic degree and classical non-linearity.
We obtained both theoretical and computational results. In particular, sufficient conditions for an S-Box to be weakly APN are presented in Propositions \ref{invariant} and \ref{derivative} and Fact \ref{sufficient}; while necessary ones can be found in Theorem \ref{converse}, Proposition \ref{degree} and Fact \ref{refine-degree}.

\section{Acknowledgements}
This research has been supported by TELSY S.p.A., MIUR ``Rientro dei cervelli'', GNSAGA of INdAM and MIUR Cofin 2008 - 
"Geo\-metria delle variet\`{a} algebriche e dei loro spazi di moduli" (Italy). 
A preliminary version of this work has been available online as arXiv:1102.3882v1 since February 17, 2011.


\begin{thebibliography}{99}

\bibitem{balanced} C. Adams and S. Tavares, \emph{The structured design of cryptographically good S-boxes},
J. Cryptology 3 (1990), no. 1, 27--41. 

\bibitem{serpent} {R. J.} Anderson and E. Biham and L.R. Knudsen,
	\emph{Serpent: A New Block Cipher Proposal},1998, p. 222--238,
	Proc. of FSE~199, LNCS,1372.


\bibitem{aubry} Y. Aubry, G. McGuire, and F. Rodier, \emph{A few more functions that are not APN infinitely often}, 
Finite fields: theory and applications, 23--31, Contemp. Math., 518, Amer. Math. Soc., Providence, RI, 2010.

\bibitem{present} A. Bogdanov and L. R. Knudsen and G. Leander and C. Paar and A. Poschmann and M. Robshaw and Y. Seurin and C. Vikkelsoe,
	\emph{PRESENT: An Ultra-Lightweight Block Cipher},
	Proc. of {CHES}~2007,
	2007,LNCS 7427, p. 450--466,

\bibitem{bracken} C. Bracken, E. Byrne, N. Markin, and G. McGuire, \emph{Fourier spectra of binomial APN functions}, 
SIAM J. Discrete Math. 23 (2009), no. 2, 596--608. 

\bibitem{quadratic} A. Canteaut, P. Charpin, and G. M. Kyureghyan, \emph{A new class of monomial bent functions}, Finite Fields Appl. 14 (2008), no. 1, 221--241.

\bibitem{CDS} A. Caranti, F. Dalla Volta, and M. Sala, \emph{On some block ciphers and imprimitive groups}, 
Appl. Algebra Engrg. Comm. Comput. 20 (2009), no. 5-6, 339--350.

\bibitem{dec} 	C. De Canni\'ere, \emph{Analysis and Design of Symmetric Encryption Algorithms}, PhD thesis, Katholieke Universiteit Leuven, 2007. 

\bibitem{Sboxes} G. Leander and A. Poschmann, \emph{On the classification of 4 bit S-boxes},  LNCS 4547, 159--176.

\bibitem{AES} 		National Institute of Standards and Technology,
       \emph{The Advanced Encryption Standard},
	(FIPS) 197, 2001

\bibitem{trapdoor} K. G. Paterson: \emph{Imprimitive permutation groups and trapdoors in iterated block ciphers}, LNCS 1636 (1999), 201--214.

\bibitem{bent} B. Preneel, W. Van Leekwijck, L. Van Linden, R. Govaerts, and J. Vandewalle:
\emph{Propagation characteristics of Boolean functions}, LNCS 473 (1991), 161--173.


\bibitem{Pul} V. Pulice: \emph{Security classification of 4-bit Boolean permutations},
Master Thesis, Univ. of Trento (2011).


\bibitem{saarinen} M. J. Saarinen, \emph{Cryptographic Analysis of All 4 x 4-Bit S-Boxes}, Proc. of SAC 2011, Toronto, Canada. 


\bibitem{permutation} W. Zhang, C.-K. Wu, and S. Li:
\emph{Construction of cryptographically important Boolean permutations},
Appl. Algebra Engrg. Comm. Comput. 15 (2004), no. 3-4, 173--177. 


\end{thebibliography}
\end{document}